\documentclass[a4paper, oneside, twocolumn, notitlepage, 10pt]{extarticle_ecoc}
\usepackage{ecoc}

\usepackage{subfig}

\usepackage{amsmath,amssymb}

\usepackage{tikz}
\usepackage{pgfplots}
\pgfplotsset{compat=1.13}

\usetikzlibrary{shapes.geometric}
\usepackage{circuitikz}

\usetikzlibrary{pgfplots.colorbrewer}
\usetikzlibrary{pgfplots.groupplots}

\usepackage{acronym}

\acrodef{ADC}{analogue-to-digital converter}
\acrodef{AWGN}{additive white Gaussian noise}
\acrodef{BtB}{back-to-back}
\acrodef{DAC}{digital-to-analogue converter}
\acrodef{DP}{dual polarisation}
\acrodef{DSP}{digital signal processing}
\acrodef{ECL}{external cavity laser}
\acrodef{EDFA}{Erbium doped fibre amplifier}
\acrodef{EGN}{extended Gaussian noise}
\acrodef{NLI}{nonlinear interference}
\acrodef{GMI}{generalised mutual information}
\acrodef{SNR}{signal-to-noise ratio}
\acrodef{SSMF}{standard single mode fibre}
\acrodef{VOA}{variable optical attenuator}

\begin{document}
\selectlanguage{british}

\title{Experimental Demonstration of Geometrically-Shaped Constellations Tailored to the Nonlinear Fibre Channel}

\author{E.~Sillekens \textsuperscript{(\textasteriskcentered)}, D.~Semrau, D.~Lavery, P.~Bayvel, R.~I.~Killey}

\maketitle

\begin{strip}
 \begin{author_descr} 
 
   Optical Networks Group, Department of Electronic and Electrical Engineering, University College London (UCL), Torrington Place, London WC1E 7JE, UK  \textsuperscript{(\textasteriskcentered)}\uline{e.sillekens@ucl.ac.uk}
 \end{author_descr}
\end{strip}

\setstretch{1.1}

\begin{strip}
  \begin{ecoc_abstract} A geometrically-shaped 256-QAM constellation, tailored to the nonlinear optical fibre channel, is experimentally demonstrated. The proposed constellation outperforms both uniform and AWGN-tailored 256-QAM, as it is designed to optimise the trade-off between shaping gain, nonlinearity and transceiver impairments.
  \end{ecoc_abstract}
\end{strip}

\section{Introduction}

Constellation shaping is a coded modulation method which has recently been widely adopted in optical fibre communications to improve spectral efficiency\cite{Cho2018TransAtlantic,Bocherer2018Joint} and flexibility to vary the modulation format. However, in the vast majority of recent demonstrations, constellations have been shaped to maximise noise tolerance in the linear \ac{AWGN} channel, and hence may be suboptimal in the presence of optical fibre nonlinearities.

The nonlinear distortion is often approximated as  \ac{NLI} noise, with the \ac{NLI} a function of the transmitted modulation format. In particular, the \ac{NLI} power can be approximated\cite{Mecozzi2012Nonlinear,Carena2014EGN} as 
\begin{equation}
	\eta_{\text{tot}}P^3 \approx\left(\eta_1+\eta_2\mathfrak{K}\right)P^3, \label{eq:NLI}
\end{equation}
with real values $\eta_1$ and $\eta_2$, total \ac{NLI} coefficient $\eta_\text{tot}$, channel launch power $P$ and the excess kurtosis 
\begin{equation}
\mathfrak{K} \triangleq \frac{ \mathbb{E}\left[|X|^4\right]}{\mathbb{E}\left[|X|^2\right]^2}-2, \label{eq:ek}
\end{equation}
of the complex constellation. Neglecting the impact of transceiver noise, the change of \ac{SNR} at optimum launch power between any two constellation A and B is given by
\begin{align}
	\frac{\text{SNR}_\text{opt,A}}{\text{SNR}_\text{opt,B}} &= \left(\frac{1+c\mathfrak{K}_\text{B}}{1+c\mathfrak{K}_\text{A}}\right)^{\frac13}, \label{eq:snrinc}
\end{align}
with $c \triangleq \frac{\eta_2}{\eta_1}$. Applying shaping to conventional QAM formats typically consists of making the constellation more Gaussian-like, increasing its excess kurtosis. Although resulting in shaping gain, high excess kurtosis reduces the optimum SNR due to \ac{NLI}, as shown by \eqref{eq:snrinc}. Consequently, an optimum balance between shaping gain and nonlinear penalty must be found to maximise system throughput.

In our previous work on probabilistic shaping\cite{Sillekens2018Simple}, we obtained a simple distribution for the nonlinear fibre channel, that outperforms the optimal distribution for the AWGN channel. The benefit of geometric shaping optimised for the nonlinear channel was also experimentally confirmed in\cite{Renner2017Experimental}. 

In this paper, we describe the design of a \textit{geometrically}-shaped constellation, specifically tailored to the nonlinear fibre model. Our proposed signal shaping is straightforward to implement, being based on a square constellation with each dimension being optimised independently. We present the results of transmission experiments demonstrating that the proposed constellation outperforms the corresponding geometrically-shaped constellation tailored to the \ac{AWGN} channel.

\section{Constellation design}

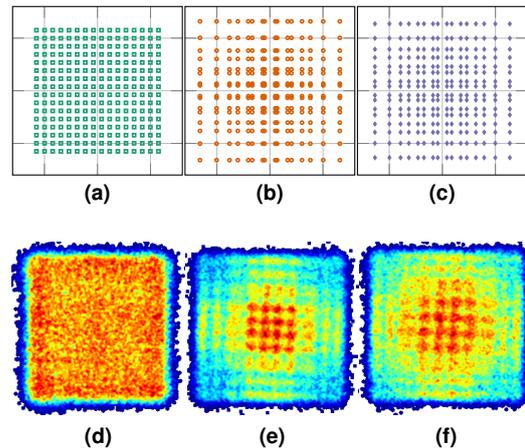
\begin{figure}[hb]
\centering
\begin{tikzpicture}
\begin{groupplot}[
  group style={
    group name=const,
    group size=3 by 1,
    horizontal sep=1pt,
    x descriptions at=edge bottom,
    y descriptions at=edge left,
  },
  height=0.5\linewidth,
  width=0.5\linewidth,
  ymin=-1.6 ,ymax=1.6 ,
   xmin=-1.6 ,xmax=1.6 ,
   grid=major,
  xticklabels={},
  yticklabels={},
every axis plot/.append style={line width=0.6pt,mark size=0.6pt},
]

\nextgroupplot

\addplot [only marks,mark=square,Dark2-A] table[col sep=comma,x=uniI,y=uniQ] {data/constellation18dBSNR.txt};

\nextgroupplot

\addplot [only marks,mark=o,Dark2-B] table[col sep=comma,x=linI,y=linQ] {data/constellation18dBSNR.txt};

\nextgroupplot

\addplot [only marks,mark=diamond,Dark2-C] table[col sep=comma,x=nlinI,y=nlinQ] {data/constellation18dBSNR.txt};

\end{groupplot}
\node[anchor=north,font=\footnotesize] at (const c1r1.south) {\bf (a)};
\node[anchor=north,font=\footnotesize] at (const c2r1.south) {\bf (b)};
\node[anchor=north,font=\footnotesize] at (const c3r1.south) {\bf (c)};
\end{tikzpicture}
\addtocounter{subfigure}{3}
\hspace*{0.2cm} \subfloat[][]{\includegraphics[width=0.30\linewidth]{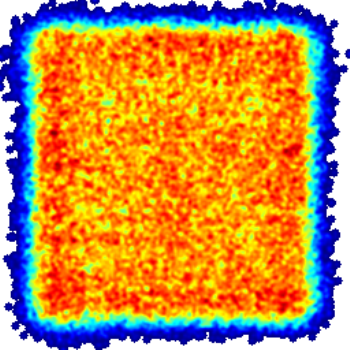}}\subfloat[][]{\includegraphics[width=0.30\linewidth]{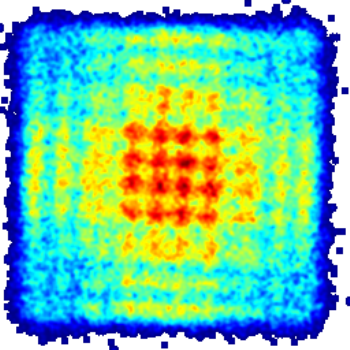}}\subfloat[][]{\includegraphics[width=0.30\linewidth]{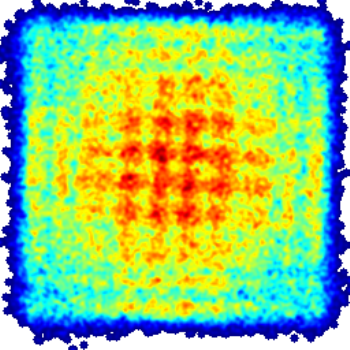}}

\caption{Geometrically-shaped constellation diagrams for 256-QAM, normalised to unit power: (a) uniform, (b) tailored to the \ac{AWGN} channel, and (c) tailored to the nonlinear fibre channel. Their respective received constellations after transmission over 160~km are shown in (d, e, f). } \label{fig:constellation}\vspace*{0.1cm}
\end{figure}

Using a quasi-Newton algorithm, the constellations are optimised for GMI. 
For the nonlinearity-tailored constellations, the SNR was changed according to Eq.~\eqref{eq:snrinc}. In-phase quadrature (IQ) independent (1D) geometric shaping was performed, as it gives four advantages: the smaller search space, its easier implementation with practical \acp{DAC}, the conservation of the square shape allows the usage of conventional blind DSP and the straightforward use of Gray coding.

Uniform square 256-QAM was used as the baseline reference format, and to obtain the system parameters of the experimental single-span 160~km long link for the central of three 35.2~GHz-spaced 35~GBd channels. Using the experimental setup, a sweep of the launch power was performed for the uniform 256-QAM. The noise figure of the amplifiers was extracted from a low launch power measurement. The nonlinear coefficient of the fibre was 1.2~W$^{-1}$km$^{-1}$. The optimum SNR for Gaussian modulation was approximately 18~dB and for the nonlinearity-tailored constellation the ratio $c=\frac{\eta_1}{\eta_2}=$~0.55 was used. The resulting constellation design is shown in Fig.~\ref{fig:constellation}.

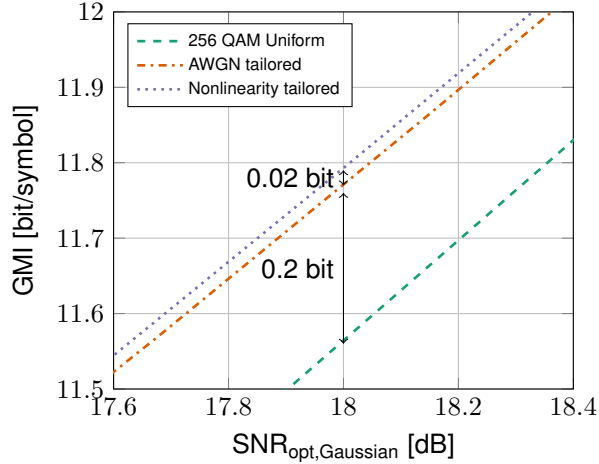
\begin{figure}[t]
\begin{tikzpicture}
\begin{axis}[
	width=\linewidth,
    xlabel={$\text{SNR}_\text{opt,Gaussian}$ [dB]},
    ylabel={GMI [bit/symbol]},
    ymin=11.5,ymax=12,
    xmin=17.6,xmax=18.4,
    grid=major,
	every axis plot/.append style={line width=1pt},
    legend cell align=left,
    legend pos=north west,
    legend style={font=\footnotesize, nodes={scale=0.8,transform shape}},
]

\addplot[Dark2-A,dashed] table[col sep=comma,x=SNRnlin,y expr=2*\thisrow{GMI}] {data/res256QAM_uni.txt};
\addlegendentry{256 QAM Uniform};

\addplot[Dark2-B,dashdotted] table[col sep=comma,x=SNRnlin,y expr=2*\thisrow{GMI}] {data/res256QAM_AWGN.txt};
\addlegendentry{AWGN tailored};

\addplot[Dark2-C,dotted] table[col sep=comma,x=SNRnlin,y expr=2*\thisrow{GMI}] {data/res256QAM_nlin.txt};
\addlegendentry{Nonlinearity tailored};

\draw[<->] (axis cs:18,11.56) -- (axis cs:18,11.76) node[midway,anchor=east] {0.2 bit};
\draw[<->] (axis cs:18,11.77) -- (axis cs:18,11.79) node[midway,anchor=east] {0.02 bit};

\end{axis}
\end{tikzpicture}
\caption{Geometrically-shaped performance.}\label{fig:design}
\end{figure}

The gains are shown in Fig.~\ref{fig:design}, where the constellation is optimised for different SNR values. At 18~dB SNR, a 0.2~bit/symbol throughput increase can be expected for the AWGN-tailored constellation, and an additional 0.02~bit/symbol increase for the nonlinearity-tailored constellation.

\section{Experimental Setup}

\begin{figure}[htb]
\centering
\includegraphics[width=\linewidth]{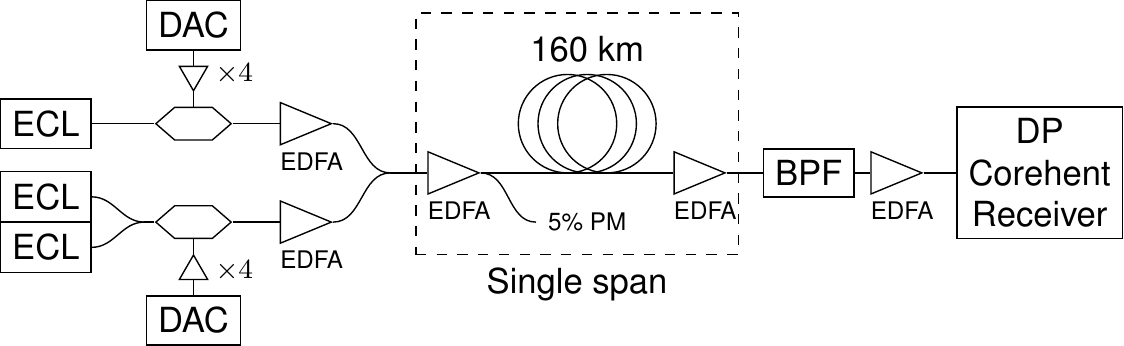}
\caption{The experimental setup, with a 3$\times$33\,GBd transmitter, single span and single channel receiver.}\label{fig:setup}
\end{figure}

The experimental setup is shown in Fig.~\ref{fig:setup}. The 3$\times$35~GBd superchannel was transmitted over a single span of 160~km of \ac{SSMF} to focus on the impact of \ac{NLI}. The launch power into the fibre span was swept to investigate the nonlinear tolerance of the constellations designed. 

The transmitter consisted of the channel under test in which a single 100 kHz-linewidth \ac{ECL} was modulated using a \ac{DP} IQ modulator. Two additional \acp{ECL} were modulated by a separate modulator to form the aggressor channels. Eight 8-bit \ac{DAC} channels operating at 87.5~GS/s, with ENOB of $\sim$4~bits at 17.5~GHz were use to generate independent channels. Both signal paths are amplified using an \ac{EDFA} and combined by a 50/50 coupler.

Before the span, an \ac{EDFA} and \ac{VOA} followed by a 5\% power tap with a power meter were used to control the launch power into the span. The 160~km span of \ac{SSMF} was followed by an \ac{EDFA}, the output of which was passed to the receiver.

The receiver had a band-pass filter and \ac{EDFA} followed by a \ac{DP} coherent receiver with a separate \ac{ECL} as local oscillator. After 160~GS/s \acp{ADC}, offline \ac{DSP} was used. After electronic dispersion compensation, a radially directed equaliser and a decision directed carrier phase estimator were used to recover the symbols. The \ac{SNR} and \ac{GMI} were extracted from the received symbols.

\section{Experimental results}

The received constellations at optimum launch powers are shown in Fig.~\ref{fig:constellation}~(d,e,f). In contrast to the uniform QAM scatter plot, the constellations tailored for the AWGN channel, shown in Fig.~\ref{fig:constellation}~(e), and for the nonlinear fibre channel, shown in Fig.~\ref{fig:constellation}~(f), set the lower energy points closer together, making these points more pronounced. The denser points should not be mistaken for probabilistically shaped constellations; all constellation points are equiprobable, the overlapping noise distributions around these points resulting in increased density of samples  within the central area.  Furthermore, it can be clearly seen that, as a result of fibre nonlinearity,the relative phase rotation between the central and outer points is higher in the AWGN tailored constellation than in the nonlinearity tailored constellation.

\begin{figure}[t]
\centering
\begin{tikzpicture}
\begin{axis}[
	width=\linewidth,
    xlabel={Launch Power [dBm]},
    ylabel={GMI [bit/symbol]},
    xmin=-2,xmax=7,
    ymin=10,ymax=12,
    grid=major,
    legend cell align=left,
    legend pos=south west,
    legend style={font=\footnotesize, xshift=1.5cm, nodes={scale=0.7,transform shape}},
    legend columns=2,
    legend image with text/.style={
        legend image code/.code={          \node[anchor=center] at (0.3cm,0cm) {#1};
        }
      },
	every axis plot/.append style={line width=1.5pt,mark size=1pt},
    ]

\addlegendimage{legend image with text=Model}
\addlegendentry{}

\addlegendimage{legend image with text=Exp.}
\addlegendentry{}

\addplot [dashed,Dark2-A] table[col sep=comma,x=LaunchPower,y expr=2*\thisrow{uniGMI}] {data/modresult3x35GBd1D.txt};
\addlegendentry{};
\addplot [mark=square,only marks,Dark2-A] table[col sep=comma,x=uniLP,y=uniGMI] {data/expresult3x35GBd1D.txt};
\addlegendentry{Uniform 256-QAM};

\addplot [dashdotted,Dark2-B] table[col sep=comma,x=LaunchPower,y expr=2*\thisrow{linGMI}] {data/modresult3x35GBd1D.txt};
\addlegendentry{};
\addplot [mark=o,only marks,Dark2-B] table[col sep=comma,x=linLP,y=linGMI] {data/expresult3x35GBd1D.txt};
\addlegendentry{AWGN tailored};

\addplot [dotted,Dark2-C] table[col sep=comma,x=LaunchPower,y expr=2*\thisrow{nlinGMI}] {data/modresult3x35GBd1D.txt};
\addlegendentry{};
\addplot [mark=diamond,only marks,Dark2-C] table[col sep=comma,x=nlinLP,y=nlinGMI] {data/expresult3x35GBd1D.txt};
\addlegendentry{Nolinearity tailored};

\draw[-,line width=1pt] (axis cs:0.5,11.6)  -- (axis cs:3.5,11.6) -- (axis cs:3.5,11.7) -- (axis cs:0.5,11.7);
\node[anchor=east] at (axis cs:0.5,11.65) {\textgreater 0.1 bit};

\end{axis}
\end{tikzpicture}
\caption{The GMI versus the launch power for all constellations. The model is shown with the lines and markers are experimental results.}\label{fig:expgmi}
\end{figure}
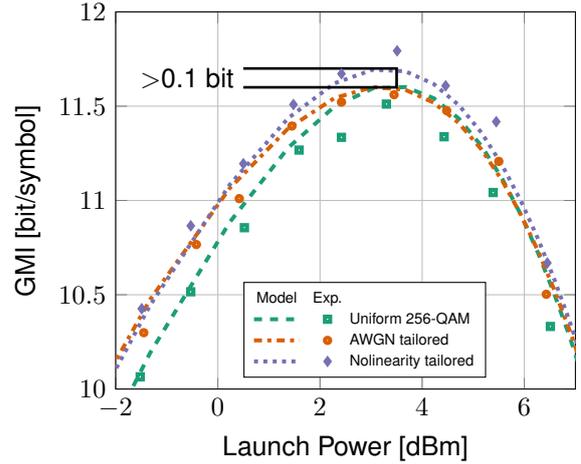

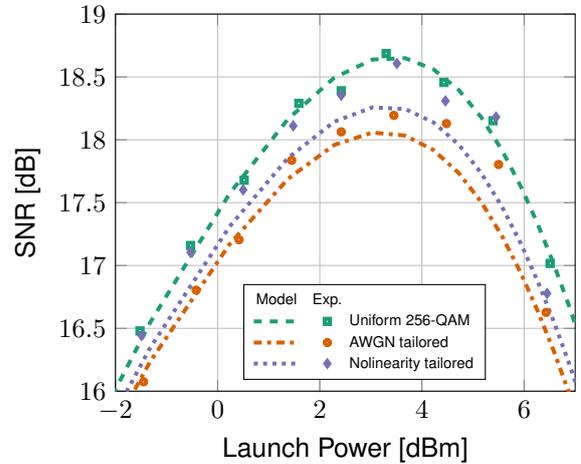
\begin{figure}[!t]
\centering
\begin{tikzpicture}
\begin{axis}[
	width=\linewidth,
    xlabel={Launch Power [dBm]},
    ylabel={SNR [dB]},
    xmin=-2,xmax=7,
    ymin=16,ymax=19,
    ytick={16,16.5,...,19},
    grid=major,
    legend cell align=left,
    legend pos=south west,
    legend style={font=\footnotesize, xshift=1.5cm, nodes={scale=0.7,transform shape}},
    legend columns=2,
    legend image with text/.style={
        legend image code/.code={          \node[anchor=center] at (0.3cm,0cm) {#1};
        }
      },
	every axis plot/.append style={line width=1.5pt,mark size=1pt},
    ]

\addlegendimage{legend image with text=Model}
\addlegendentry{}

\addlegendimage{legend image with text=Exp.}
\addlegendentry{}

\addplot [dashed,Dark2-A] table[col sep=comma,x=LaunchPower,y=uniSNR] {data/modresult3x35GBd1D.txt};
\addlegendentry{};
\addplot [mark=square,only marks,Dark2-A] table[col sep=comma,x=uniLP,y=uniSNR] {data/expresult3x35GBd1D.txt};
\addlegendentry{Uniform 256-QAM};

\addplot [dashdotted,Dark2-B] table[col sep=comma,x=LaunchPower,y=linSNR] {data/modresult3x35GBd1D.txt};
\addlegendentry{};
\addplot [mark=o,only marks,Dark2-B] ttable[col sep=comma,x=linLP,y=linSNR] {data/expresult3x35GBd1D.txt};
\addlegendentry{AWGN tailored};

\addplot [dotted,Dark2-C] table[col sep=comma,x=LaunchPower,y=nlinSNR] {data/modresult3x35GBd1D.txt};
\addlegendentry{};
\addplot [mark=diamond,only marks,Dark2-C] table[col sep=comma,x=nlinLP,y=nlinSNR] {data/expresult3x35GBd1D.txt};
\addlegendentry{Nolinearity tailored};

\end{axis}
\end{tikzpicture}
\caption{The SNR versus the launch power for all constellations. The model is shown with the lines and markers are experimental results.} \label{fig:expsnr}
\end{figure}

The experimentally-measured GMIs and SNRs are shown in Fig.~\ref{fig:expgmi} and Fig.~\ref{fig:expsnr}, respectively. The markers are experimentally obtained and the lines are from the model. The \ac{BtB} \ac{SNR} and $\eta_\text{tot}$ derived from the model are shown in Tab.~\ref{tab:btb_snr}.

\begin{table}[h]
   \centering
\caption{Back-to-back SNR and $\eta_\text{tot}$ for the constellations.} \label{tab:btb_snr}
\begin{tabular}{|l|c|c|}
         \hline    & BtB SNR [dB] & $\eta_\text{tot}$ [dB] \\
         \hline	Uniform  				& 22.78 	& 27.61 \\
         \hline  AWGN tailored  		& 21.63 	& 28.23 \\
         \hline  Nonlinearity tailored  	& 22.01 	& 28.08 \\
         \hline
\end{tabular}
\end{table}

The uniform QAM exhibited the highest \ac{BtB} \ac{SNR} and the lowest $\eta_\text{tot}$, resulting in the highest \ac{SNR} at optimal launch power. However, this modulation format has no shaping gain and consequently has the lowest \ac{GMI} of the three constellation formats evaluated, with an optimum value of 11.6~bit/symbol.

The \ac{AWGN} tailored format has the lowest \ac{SNR} across all launch powers, but outperforms uniform QAM (i.e. higher GMI) at launch powers below the optimal. At higher launch powers the model does not predict any gains, while the experimentally observed performance was marginally higher.

Due to lower excess kurtosis, the nonlinearity tailored constellation has a $\eta_\text{tot}$ lower than the \ac{AWGN} tailored constellation  as predicted by Eq.~\eqref{eq:NLI}. Additionally, the lower excess kurtosis results in a higher \ac{BtB} \ac{SNR} because of the lighter tailed distributions of the constellation and resulting reduced quantisation noise. The nonlinearity tailored constellation offers a trade-off between shaping gain and the impact of the nonlinear interference. It achieved a GMI of 11.7~bit/symbol after transmission over the 160~km link, a \textgreater~0.1~bit/symbol increase over the other two constellations.

\section{Conclusions}

We experimentally demonstrated a nonlinearity tailored distribution that outperforms the distribution tailored to the AWGN channel by \textgreater~0.1~bit/symbol, a result of a trade-off between fibre nonlinearity, transceiver limitations and shaping gain. The work indicates that all three aspects should be taken into consideration to maximise the system performance. However, empirically we have shown that, by reducing the excess kurtosis of a modulation format, the tolerance to both fibre nonlinearity and transceiver limitations can be simultaneously improved.

\section{Acknowledgements}
{
This work is financially supported by the EPSRC UNLOC, INSIGHT and and TRANSNET grants through a PhD studentship to Eric Sillekens (grant EP/M507970/1) and Xtera Communications Inc.
}

\bibliographystyle{IEEEtran}
\begin{spacing}{1}

\end{spacing}
\vspace{-4mm}

\end{document}